\input harvmac
\input epsf
\input amssym

\tolerance=10000



%
\def\omit#1{}

\def\coeff#1#2{\relax{\textstyle {#1 \over #2}}\displaystyle}
\def\half{{1 \over 2}}

\def\oneone{\rlap 1\mkern4mu{\rm l}}

\def\cN{{\cal N}} 
 
\def\cR{{\cal R}}

\def\ie{{\it i.e.}}
\def\bfone{\relax{\rm 1\kern-.35em 1}}

%
\def\us{\bf}

\def\IR{\Bbb{R}}
\def\nup#1({Nucl.\ Phys.\ $\us {B#1}$\ (}
\def\plt#1({Phys.\ Lett.\ $\us  {#1B}$\ (}
\def\plt#1({Phys.\ Lett.\ $\us  {#1B}$\ (}
\def\cmp#1({Comm.\ Math.\ Phys.\ $\us  {#1}$\ (}
\def\prp#1({Phys.\ Rep.\ $\us  {#1}$\ (}
\def\prl#1({Phys.\ Rev.\ Lett.\ $\us  {#1}$\ (}
\def\prv#1({Phys.\ Rev.\ $\us  {#1}$\ (}
\def\mpl#1({Mod.\ Phys.\ Let.\ $\us  {A#1}$\ (}
\def\ijmp#1({Int.\ J.\ Mod.\ Phys.\ $\us{A#1}$\ (}
\def\atmp#1({Adv.\ Theor.\ Math.\ Phys.\ $\bf {#1}$\ (}
\def\cqg#1({Class.\ Quant.\ Grav.\ $\bf {#1}$\ (}
\def\jag#1({Jour.\ Alg.\ Geom.\ $\us {#1}$\ (}
\def\jhep#1({JHEP $\bf {#1}$\ (}

%



%
%
%
\lref\MaldacenaYY{
J.~M.~Maldacena and C.~Nunez,
``Towards the large N limit of pure N = 1 super Yang Mills,''
Phys.\ Rev.\ Lett.\  {\bf 86}, 588 (2001)
[arXiv:hep-th/0008001].
}
%
\lref\GauntlettPS{
J.~P.~Gauntlett, N.~Kim, D.~Martelli and D.~Waldram,
``Wrapped fivebranes and N = 2 super Yang-Mills theory,''
Phys.\ Rev.\ D {\bf 64}, 106008 (2001)
[arXiv:hep-th/0106117].
}
%
%
\lref\GauntlettSC{
J.~P.~Gauntlett, D.~Martelli, S.~Pakis and D.~Waldram,
``G-structures and wrapped NS5-branes,''
arXiv:hep-th/0205050.
}
%
%
\lref\GauntlettCY{
J.~P.~Gauntlett, D.~Martelli and D.~Waldram,
``Superstrings with intrinsic torsion,''
arXiv:hep-th/0302158.
}
%
\lref\GauntlettRV{
J.~P.~Gauntlett, N.~Kim, S.~Pakis and D.~Waldram,
``M-theory solutions with AdS factors,''
Class.\ Quant.\ Grav.\  {\bf 19}, 3927 (2002)
[arXiv:hep-th/0202184].
}
%
\lref\GauntlettNW{
J.~P.~Gauntlett, J.~B.~Gutowski, C.~M.~Hull, S.~Pakis and H.~S.~Reall,
``All supersymmetric solutions of minimal supergravity in five  dimensions,''
arXiv:hep-th/0209114.
}
%
%
\lref\GauntlettFZ{
J.~P.~Gauntlett and S.~Pakis,
``The geometry of D = 11 Killing spinors,''
arXiv:hep-th/0212008.
}
%
%
\lref\TodPM{
K.~P.~Tod,
``All Metrics Admitting Supercovariantly Constant Spinors,''
Phys.\ Lett.\ B {\bf 121}, 241 (1983).
}
%
\lref\TodJF{
K.~P.~Tod,
``More On Supercovariantly Constant Spinors,''
Class.\ Quant.\ Grav.\  {\bf 12}, 1801 (1995).
}
\lref\PopeBD{
C.~N.~Pope and N.~P.~Warner,
``An SU(4) Invariant Compactification Of D = 11 Supergravity On A Stretched Seven Sphere,''
Phys.\ Lett.\ B {\bf 150}, 352 (1985).
}
%
%
\lref\GowdigereUK{
C.~N.~Gowdigere and N.~P.~Warner,
``Flowing with eight supersymmetries in M-theory and F-theory,''
arXiv:hep-th/0212190.
}
%
\lref\PopeJP{
C.~N.~Pope and N.~P.~Warner,
``A dielectric flow solution with maximal supersymmetry,''
arXiv:hep-th/0304132.
}
%
\lref\GibbonsZT{
G.~W.~Gibbons and S.~W.~Hawking,
``Gravitational Multi - Instantons,''
Phys.\ Lett.\ B {\bf 78}, 430 (1978).
}
%
\lref\CorradoNV{
R.~Corrado, K.~Pilch and N.~P.~Warner,
``An N = 2 supersymmetric membrane flow,''
Nucl.\ Phys.\ B {\bf 629}, 74 (2002)
[arXiv:hep-th/0107220].
}
%
\lref\KhavaevYG{
A.~Khavaev and N.~P.~Warner,
``An N = 1 supersymmetric Coulomb flow in IIB supergravity,''
Phys.\ Lett.\ B {\bf 522}, 181 (2001)
[arXiv:hep-th/0106032].
}
%
\lref\PilchFU{
K.~Pilch and N.~P.~Warner,
``N = 1 supersymmetric renormalization group flows from IIB supergravity,''
Adv.\ Theor.\ Math.\ Phys.\  {\bf 4}, 627 (2002)
[arXiv:hep-th/0006066].
}
%
\lref\PilchUE{
K.~Pilch and N.~P.~Warner,
``N = 2 supersymmetric RG flows and the IIB dilaton,''
Nucl.\ Phys.\ B {\bf 594}, 209 (2001)
[arXiv:hep-th/0004063].
}
%
\lref\PilchEJ{
K.~Pilch and N.~P.~Warner,
``A new supersymmetric compactification of chiral IIB supergravity,''
Phys.\ Lett.\ B {\bf 487}, 22 (2000)
[arXiv:hep-th/0002192].
}
\lref\GNW{C.~Gowdigere, D.~Nemeschansky and N.~P.~Warner, to appear}
\lref\PWIIB{K.~Pilch and N.~P.~Warner,  
``Generalizing the $N=2$ supersymmetric RG flow  solution of IIB supergravity,''
USC-03/05,  to appear }
%
\Title{ \vbox{ \hbox{USC-03/02} 
\hbox{\tt hep-th/0306097} }} {\vbox{\vskip -1.0cm
\centerline{\hbox
{Supersymmetric Solutions with Fluxes }}
\vskip .5cm
\centerline{\hbox { from Algebraic Killing Spinors }}
\vskip 8 pt
\centerline{
\hbox{}}}}
\vskip -0.9 cm
\centerline{Chethan N.\ Gowdigere, Dennis Nemeschansky and  Nicholas P.\ Warner} 
\bigskip
\centerline{ {\it Department of Physics and Astronomy}}  
\centerline{{\it University of Southern California}} 
\centerline{{\it Los Angeles, CA
90089-0484, USA}} 

\vskip 0.8cm
\centerline{{\bf Abstract}}
\medskip
We give a general framework for constructing supersymmetric  
solutions in the presence of non-trivial fluxes of tensor gauge fields.
This technique involves making a general Ansatz for the metric and then
defining the Killing spinors in terms of very simple projectors on the spinor
fields.  These projectors and, through them, the spinors, are determined
{\it algebraically} in terms of the metric Ansatz.  The Killing spinor equations
then fix the tensor gauge fields algebraically, and, with the Bianchi identities,
provide a  system of equations for all the metric functions.
We illustrate this by constructing an infinite family of massive flows that
preserve eight supersymmetries in $M$-theory.  This family constitutes all the
radially symmetric Coulomb branch flows of the softly broken, large $N$ scalar-fermion
theory on $M2$-branes.  We reduce the problem to the solution of a single,
non-linear partial differential equation in two variables.  This
equation governs the flow of the fermion mass, and the function that
solves it then generates the entire $M$-theory solution algebraically in terms of 
the function and its first derivatives.
While the governing equation is non-linear, it has a very simple perturbation theory
from which one can see how the Coulomb branch is encoded.
\vskip .1in
\Date{\sl {June, 2003}}

\parskip=4pt plus 15pt minus 1pt
\baselineskip=15pt plus 2pt minus 1pt

\newsec{Introduction}

It has been an outstanding open problem in AdS/CFT to find
a relatively simple characterization of {\it supersymmetric } backgrounds that involve
non-trivial fluxes.   Such solutions are particularly important in the study
of holographic RG flows in supergravity:  First, there is always the flux generated
by the branes upon which the holographic field theory lies, and  then one often 
wants to add further fluxes that are typically holographically dual to fermion
mass terms on the  brane.   Such ``multiple flux'' solutions thus lie at the heart 
of the holographic study of softly broken supersymmetry.  

There are two basic methods of generating such solutions:  One can either
work with the appropriate supergravity theory in ten or eleven dimensions,
or one can work with the appropriate gauged supergravity in lower dimensions.
While some beautiful results have been obtained using the latter approach, it is
ultimately limited because a gauged supergravity theory truncates the fields of the
higher-dimensional theory to some especially simple sub-sector of  ``lowest
harmonics''.  The resulting solutions thus involve very symmetric fluxes with very
smooth brane distributions.  The  huge advantage of gauged supergravity is 
that it produces an extremely simple description of flows whose 
higher dimensional analogues, or ``lifts''  can be extremely complicated.
One version of the problem that we will address in this paper is how to 
extract a general geometric principle that may be used to construct 
multi-flux solutions directly in ten or eleven dimensions, and yet embody 
the remarkable simplicity apparent in the gauged supergravity description.

The second method of generating supersymmetric flux solutions is to work
directly in ten or eleven dimensions.  There have been some beautiful
results, like the ``harmonic principle'' for intersecting branes.   In holographic
field theory this leads to, amongst other things, a very nice picture of
the Coulomb branch of a supersymmetric theory.  More generally, one
would like to study flows that not only involve Coulomb branch flows, but
also involve softly broken supersymmetry.    The difficulty lies in attempting
to classify such solutions with ``multiple fluxes'' that lie in different
directions with respect to the underlying branes.  
Much important work has been done of work on this,
particularly for solutions  that involve wrapped $NS$ $5$-branes
 (see, for example, \refs{ \MaldacenaYY, \GauntlettPS, \GauntlettSC , 
 \GauntlettRV, \GauntlettCY}).    There was also some important early work on
four-dimensional Einstein-Maxwell solutions \refs{\TodPM, \TodJF}.
The general philosophy of the more recent papers on this subject 
has been to try to exploit the ideas of $G$-structures (see for example, 
\refs{\GauntlettNW,  \GauntlettRV, \GauntlettCY}).
This approach is very promising, but the concept of a $G$-structure is
rather a general one, and thus far,  the new results have been largely restricted to
low dimensions or to the consideration of a single background flux, and most
particularly to a flux that my be thought of as a torsion.    

In this paper we will present an approach to generating whole new
families of solutions with multiple fluxes.  In particular, we believe that
our approach will enable one to obtain the most general holographic analogues
of the Donagi-Witten ($\cN=2^*$) flows to an arbitrary point on the Coulomb
moduli space.  While we do not solve this particular problem here, we do
solve an $M$-theory analogue:  We find an infinite family of $M$-theory
flow solutions with eight supersymmetries.  This new family is   
obtained from a solution to a second order PDE in two variables, and it represents
a generalization of the flow of \GowdigereUK\ to a spherically symmetric
Coulomb distribution of  $\cN=4^*$ flows of the $\cN=8$ scalar-fermion
theory in $(2+1)$-dimensions.  The fact that we can find such a solution 
represents progress in what has been a technically complicated subject, however
we think that the method by which it is obtained will admit many interesting
variations and generalizations.  Indeed, an analysis of a class of
$\cN=2^*$ flows in IIB supergravity will appear in \PWIIB.

Our approach is very much in the spirit of the ideas of $G$-structures, however 
rather than constructing all the differential forms associated with the 
Killing spinor, we make a rather general Ansatz for the metric, and then we
make an {\it algebraic}  Ansatz for the Killing spinors.  To be more precise, 
we will work in eleven dimensions, and thus a Killing spinor will mean a solution of 
the equation:
\eqn\gravvar{\delta \psi_\mu ~\equiv~ \nabla_\mu \, \epsilon ~+~ \coeff{1}{144}\,
\Big({\Gamma_\mu}^{\nu \rho \lambda \sigma} ~-~ 8\, \delta_\mu^\nu  \, 
\Gamma^{\rho \lambda \sigma} \Big)\, F_{\nu \rho \lambda \sigma} ~=~ 0 \,.}
We will require that the space of Killing spinors be defined by two very simple 
projection operators, $\Pi_0$ and $\Pi_1$, on the spinor space.   Each projector 
reduces the number of spinor components by half, and so the two projectors reduce the 
32-component spinor to eight components.  These remaining eight spinors, $\epsilon^j$,
$j=1,\dots,8$  are then required to be supersymmetries.  
The form of the projector, $\Pi_1$,  is elementary and it is naturally
motivated in terms of the moduli space of  brane probes.  The other projector, $\Pi_0$,
is a non-trivial deformation of the usual Dirichlet projector parallel to the
$M2$-brane.  The key to understanding this projector lies in the fact that the
spinor bilinears:
\eqn\Kvecs{K_{(ij)}^\mu ~\equiv~ \bar \epsilon^i \, \Gamma^\mu \, \epsilon^j \,,}
are necessarily Killing vectors in $M$-theory.   For a harmonic $M2$-brane
distribution, equation \Kvecs\ will only give Killing vectors, $K^\mu$, parallel to the 
$M2$-brane.   The deformation is constructed so that \Kvecs\ gives rise to 
another ``internal'' Killing vector, $L^\mu$, transverse to the original $M2$-brane 
distribution.   This deformation involves an,  {\it a priori}, arbitrary function.
However, this function is fixed in terms of the metric coefficients
by requiring that $L^\mu$ is indeed a Killing
vector (and not a functional multiple of a Killing vector).   Moreover, requiring that
the vectors, $K^\mu$,  parallel to the branes are Killing vectors completely
determines the normalization of the Killing spinors in terms of the metric.  Thus the 
Killing spinor Ansatz is completely determined in terms of {\it algebraic}
combinations of functions that appear in the metric Ansatz.  
Since the Killing spinor equation is a first order
differential equation, the Killing spinors would normally involve first derivatives
of metric coefficients.  Our approach thus produces a rather special class
of ``algebraic Killing spinors''.   

Our result generalizes the well known result
about Killing spinors for harmonic distributions  of branes: 
\eqn\harmKsp{\epsilon ~=~ H^{-\alpha} \, \epsilon_0 \,,}
in which $\epsilon_0$ is a constant spinor, $\alpha$ is a rational
number (${1 \over 6}$ for $M2$-branes) and $H$ is the harmonic function
that appears in the metric.  Such harmonic solutions 
have 16 supersymmetries in maximal supergravity.   The formula  \harmKsp\
can be deduced from the form of the metric and requiring that 
\Kvecs\ yields Killing vectors.    Our Killing spinors 
will be characterized by two metric functions.  The second function 
arises in the further reduction of the supersymmetry.  

We believe that our method can easily be adapted to study 
softly broken theories in which there is a non-trivial space of moduli, such as a
a Coulomb branch.   For example, there will presumably be whole classes of solutions 
where that involve another projector, $\Pi_2$, that reduces the number of symmetries to four.   
The projector, $\Pi_2$, could be obtained from the reduction of the supersymmetry on
the space of moduli.

Having made a metric Ansatz and fixed the Killing spinors there remains the problem of
reconstructing  the background tensor gauge field.  This turns out to be very straightforward:
The Killing spinor equation \gravvar\ can be used to determine the Maxwell tensor 
algebraically in terms of the metric and Killing spinor.  The interesting differential
equations will then emerge from the Bianchi identities.

Alternatively, one can use some of the $G$-structure to obtain the 
tensor gauge fields.  That is, one defines  forms   from bilinears of Killing spinors:
\eqn\Forms{\Omega^{(ij)}_{\mu_1\mu_2 \dots \mu_p} ~\equiv~ \bar \epsilon^i 
\, \Gamma_{\mu_1\mu_2 \dots \mu_p}\, \epsilon^j \,.}
These forms satisfy first order differential equations that may be derived from \gravvar.  
These equations have been catalogued in \GauntlettFZ.  We 
can use one or two of the simpler equations to deduce almost all 
the components of the background $3$-form potential, $A^{(3)}_{\mu \nu \rho}$.
Specifically,  we can show that if  $K_{(ij)}^\rho$ is a Killing vector given by
\Kvecs\ then there is a gauge in which:
\eqn\AdotK{  \Omega^{(ij)}_{\mu \nu} ~\sim~ A^{(3)}_{\mu \nu \rho} \, K_{(ij)}^\rho\,.}
Thus we may obtain most of the components of the tensor gauge field.  Indeed, in 
the example that we consider here,  this gives us all but one component of
$ A^{(3)}$.  Again notice that these components are algebraic combinations of the metric functions.

Either way, it is very straightforward to obtain an expression for   $F_{\nu \rho \lambda \sigma}$
in terms of derivatives of metric functions.  In the specific example 
considered in this paper,  the Bianchi identities then reduce to a single second order, 
{\it non-linear} PDE:
\eqn\basicPDE{  {1 \over u^3}\, {\del \over \del u} \Big( \,u^3 {\del \over \del u}
\Big( {1 \over u^2} \,  g(u,v)  \Big) \Big) ~+~  {1 \over v}  \,
{\del  \over \del v} \Big(  g(u,v)  \, {1 \over  v}\,  {\del \over \del v}\,
\Big( {v^2  \over u^2}  \,  g(u,v)  \Big) \Big) ~=~ 0\,.}
While we do not know if this PDE is explicitly solvable in general, it does have a very
simple perturbation theory in which the solution at $n^{th}$ order involves
solving a simple linear PDE  with a source that is a quadratic form in
the lower-order solutions and their derivatives.  The zeroeth order solution is a constant,
and the first order ``seed'' is {\it any} solution of the homogeneous linear equation:
\eqn\linearPDE{   {1 \over u}\, {\del \over \del u} \Big( \,u^3 {\del \over \del u}
\Big( {1 \over u^2} \,  g(u,v)  \Big) \Big) ~+~  {1 \over v}  \,
{\del  \over \del v} \Big(    \, {1 \over  v^3 }\,  {\del \over \del v}\,
\big( v^4  \,  g(u,v)  \big) \Big) ~=~ 0 \,.}
Thus we see that there is an infinite family of solutions to the non-linear PDE
generated by the family of solutions of the linearized PDE.

There is one potential danger in our procedure:  Solving the Killing spinor
equation does {\it not} necessarily  guarantee a solution to the full set of equations of
motion of the theory.   This seems to violate some basic preconceptions about 
supersymmetry, but harmonic distributions of branes illustrate this point.
One can easily check that the Ansatz \harmKsp\ satisfies \gravvar\ for the
appropriate ``harmonic'' Ansatz for the metric and tensor gauge field.  However
the Killing spinor equation works for any arbitrary function $H$:  The harmonic 
condition on $H$ {\it only} comes from imposing the equations of motion\foot{We
are grateful to J. Gomis for pointing this out to us.}.  
The supergravity preconception arises from the fact that the commutator
of two supersymmetries generates equations of motion, and so it seems 
that one cannot solve \gravvar\ without solving the equations of motion.  However,
the commutator of two supersymmetries does not necessarily generate {\it all} equations 
of motion.  For the harmonic brane configuration it generates a combination
of the Einstein and Maxwell equations from which the Laplacian on $H$ cancels.
Thus there is a danger that solving the equations of motion might impose further 
conditions on our solution and perhaps render it trivial.  This does not happen in
our example:  We have explicitly checked that the second order, non-linear 
PDE is both necessary and sufficient to solve all the equations of motion.  
While we verified every equation of motion, it turns out that the 
procedure can be significantly simplified:  There is a very nice discussion of 
``sufficiency''   in section 2 of \GauntlettFZ, where it is shown, very generally, that once 
one has satisfied the supersymmetry variations, one only needs to check, at most, one 
of the Einstein equations and only a subset of the Maxwell  equations.     
We also suspect that any remaining   ``insufficiency'' of solving  \gravvar\ is a 
pathology of the special structure of the ``harmonic'' brane Ansatz, and that 
for any sufficiently complicated solution (like ours), solving \gravvar\ will be sufficient 
to solve all the equations of motion.

In section 2 of this paper we will describe our basic approach to generating families of
supersymmetric flux solutions, while in section 3 we will consider a detailed example 
the generalizes the flow of \GowdigereUK.  Section 4 contains some remarks
about the geometry transverse  to the branes, and section 5 contains some
final comments.

\newsec{The Ansatz: Generalities}

Our purpose here is to outline the aspects of our Ansatz that should generalize 
readily to other settings.  While our comments will be made primarily for $M$-theory,
they should be readily applicable to any supergravity theory.

Our $M$-theory conventions are those of \PopeBD.  Our metric is ``mostly plus,''
and we will take the gamma-matrices to be
\eqn\gammamats{\eqalign{& \Gamma_1 ~=~  -i \, \Sigma_2 \otimes 
\gamma_9 \,, \quad
\Gamma_2~=~  \Sigma_1 \otimes \gamma_9 \,, \quad 
\Gamma_3~=~   \Sigma_3 \otimes \gamma_9 \,, \cr &
\Gamma_{j+3} ~=~  {\oneone}_{2 \times 2} \otimes \gamma_j \,, 
\quad j=1,\dots,8 \,,}}
where the $\Sigma_a$ are the Pauli spin matrices, $\oneone$ is
the Identity matrix, and  the $\gamma_j$ are real, symmetric $SO(8)$
gamma matrices.  As a result, the $\Gamma_j$ are all real, with 
$\Gamma_1$ skew-symmetric and $\Gamma_j$ symmetric for $j>2$.  One also
has:
$$
\Gamma^{1 \cdots\cdots 11} ~=~ \oneone \,,
$$
where $\oneone$ will henceforth denote the $32 \times 32$ identity matrix.
The gravitino variation will be as in \gravvar.   With these conventions,
sign choices and normalizations,  the equations of motion are:
\eqn\eqnmot{\eqalign{ R_{\mu \nu} ~+~ R \, g_{\mu \nu}  ~=~&  \coeff{1}{3}\, 
F_{\mu \rho \lambda \sigma}\, F_\nu{}^{ \rho \lambda \sigma}\,,\cr 
\nabla_\nu F^{\mu \nu \rho  \sigma} ~=~& -  \coeff{1}{ 576} \, \varepsilon^{\nu \rho  \sigma
\lambda_1\lambda_2
\lambda_3 \lambda_4 \tau_1\tau_2  \tau_3 \tau_4} \,F_{ \lambda_1  \lambda_2
 \lambda_3  \lambda_4} \, F_{ \tau_1\tau_2 \tau_3 \tau_4} \,.}}

\subsec{The harmonic  brane solutions}

It is first worth recalling the situation for a harmonic distribution of $M2$ branes.
The metric is given by:
\eqn\harmmet{ ds_{11}^2 ~=~ H^{-2/3}  \, (   -dt^2 +   dx_1^2 + 
dx_2^2)    ~+~ H^{1/3}  \,   \bigg(   \sum_{j=1}^8 \, dy^j \, dy^j \bigg)\,,}
and the Killing spinors are given by 
\eqn\MtwoharmKsp{\epsilon ~=~ H^{-{1 \over 6}} \, \epsilon_0 \,,}
where $\epsilon_0$ is a constant spinor satisfying the projection condition
\eqn\DirProj{  \big(\oneone  ~-~ \Gamma^{123} \big) \, \epsilon_0 ~=~0 \,. }
Consider a Killing vector of the form:
\eqn\sampleKvec{K_{(ij)}^\mu ~\equiv~ \bar \epsilon^i \, \Gamma^\mu \, \epsilon^j \,,}
Observe that $\Gamma^{123}$ is hermitian (indeed, it is real and symmetric)
and anti-commutes with $\Gamma^\mu$ for $\mu =4,5,\dots, 11$.  Inserting 
$\Gamma^{123}$ in front of $\epsilon^j$ in \sampleKvec\ still yields
$K_{(ij)}^\mu$ because of the projection condition \DirProj.  Now commute it
through $\Gamma^\mu$, and use its action on $\bar \epsilon^i$ (remembering that
the Dirac conjugate contains $\Gamma^1$), and one gets 
$- K_{(ij)}^\mu$ for $\mu =4,5,\dots, 11$.  We thus learn that $ K_{(ij)}^\mu \equiv 0$
for $\mu =4,5,\dots, 11$, and hence any Killing vectors generated by \sampleKvec\
must be parallel to the brane.

Since we will use the foregoing kind of argument several times,  will refer to it
 as ``the standard projector argument.''

\subsec{The generalized Ansatz}

In our Ansatz we will deform the  projector in \DirProj\ to:
\eqn\deformProj{\Pi_0 ~\equiv ~ \half\, \big(\oneone  ~-~ p_1 \,  \Gamma^{123} +
 p_2 \,  \Gamma^{*} \big)\,,}
where $p_1$ and $p_2$ are functions, and $\Gamma^{*}$ is a product of gamma-matrices 
satisfying $( \Gamma^{*})^2 = \oneone$.  For  $\Pi_0 $ to be a projector one must have:
\eqn\quadreln{ p_1^2 ~+~ p_2^2 ~=~ 1 \,.}
One can obviously generalize \deformProj\ to involve several functions
and sums of products of gamma matrices (see, for example \PopeJP).  However,
\deformProj\ will be sufficient for our purposes here.
The obvious issue is  the gamma-matrices that make up  $\Gamma^{*}$. 
We will return to this after fixing the other projector,  
but we note that Lorentz invariance on the brane means that
$\Gamma^{*}$ is made out of the eight gamma-matrices\foot{It could involve the 
product $\Gamma^{123}$, but
then we would multiply \deformProj\ by $\Gamma^{123}$ to get rid of
this product in $\Gamma^{*}$.}:  $\Gamma^4,\dots
\Gamma^{11}$.  Moreover,  reality of the Killing spinor and the requirement
that $( \Gamma^{*})^2 = \oneone$ means that  $\Gamma^{*}$ must involve
either one, four, five, or seven of the gamma-matrices $\Gamma^4,\dots
\Gamma^{11}$.   

Since we wish to focus on flows with eight supersymmetries, we will need a
second projection matrix, which we will take to be of the form:
\eqn\secondProj{\Pi_1 ~\equiv~ \half\, \big(\oneone  ~+~  \widehat \Gamma \big)\,,}
where $ \widehat \Gamma$ is also a product of gamma-matrices 
with $( \widehat \Gamma )^2 = \oneone$.  The choice of $\widehat \Gamma$ is
fixed by the projection condition on the supersymmetries when they
are restricted to the moduli space of brane probes.  In this instance we will
take $ \widehat \Gamma$ to be the product of all gamma-matrices parallel
to the brane-probe moduli space.  For flows with eight supersymmetries in M-theory,
the moduli space will be a four-dimensional hyper-K\"ahler manifold and choosing
$\Pi_1$ in this manner will impose the proper ``half-flat''   chirality condition on the 
Killing spinors.   More generally, for four supersymmetries one will get a K\"ahler moduli 
space and one will need to impose further projection conditions.   
 
We will denote the coordinates parallel to the moduli-space by $x^5, x^9, x^{10},  x^{11}$,
and hence we will take:
\eqn\Projone{\Pi_1 ~\equiv~\half\, \big(\oneone  ~+~   \Gamma^{59\,10\,11} \big)\,.}

We have now completely fixed the space of Killing spinors, $\epsilon^j$, $j=1,\dots,8$.
The standard projector argument  using $\widehat \Gamma$ shows that $K_{(ij)}^\mu$
can {\it never} be parallel to the moduli space.  
Now recall that solutions with eight supersymmetries in eleven dimensions have
an $SU(2) \times  SU(2)$ $\cR$-symmetry.  For the flows considered here, one of these 
$SU(2)$'s acts on the family of complex structures in the hyper-K\"ahler moduli space, while
the second $SU(2)$ acts on the geometry transverse to both the brane and moduli space.
Additional isometries of the metric transverse to the brane amount to additional
global symmetries of the field theory on the brane.
Thus the second $SU(2)$ of the $\cR$ symmetry  will act in the $x^4, x^6, x^{7},  x^{8}$ 
directions.  Indeed, we will assume that the metric involves the right-invariant $1$-forms, 
$\sigma_j$, of $SU(2)$ and that the frames $e^6, e^{7},  e^{8}$ will be proportional to the 
$\sigma_1,\sigma_2,\sigma_3$ respectively.

To fix the deformation of $\Pi_0$ we first require that it be compatible
(commute) with $\Pi_1$.  This means that it must contain an even number
of the gamma-matrices $\Gamma^5,\Gamma^9,\Gamma^{10},  \Gamma^{11}$.
More fundamentally, one must decide how the Killing
vectors of \sampleKvec\ relate to the $\cR$-symmetries and to any additional
global symmetries of the field theory.   Here we are going to require
\sampleKvec\  generates only one Killing vector {\it transverse} to the brane.
In particular, we will require that \sampleKvec\ generate none of
the $\cR$-symmetries, but instead generate a single, $U(1)$, global symmetry
of the underlying field theory.  This symmetry will  be taken to be a translation parallel to the 
$x^8$-coordinate (and it will thus act by rotating $\sigma_1$ and 
$\sigma_2$ into one another).  Given that anything not forbidden will appear 
in \sampleKvec, we must construct the projector $\Pi_0$ so that it forbids the
proper things via the standard projector argument.  Since the M-theory
solution will generically depend on all the non-symmetry coordinates ($x^4$,
$x^5, x^9, x^{10},  x^{11}$) these will not generally be Killing directions, and hence 
must be forbidden from appearing in \sampleKvec.  This means that
$\Gamma^*$  must contain  $\Gamma^4$.
Similarly, since we required \sampleKvec\ not to yield an $\cR$-symmetry, 
$\Gamma^*$  must contain  $\Gamma^{67}$.  The matrix $\Gamma^*$ must then
be  completed by any two of  $\Gamma^5,\Gamma^9,\Gamma^{10},  \Gamma^{11}$.
Since, as yet,  there is no special meaning to any of the coordinate labels 
on the moduli space, we will take:
\eqn\gammastar{ \Gamma^* ~\equiv~ \Gamma^{456711}  \quad \Rightarrow 
\quad \Pi_0 ~\equiv~ \half\,\big(\oneone  ~-~ p_1 \,  \Gamma^{123} +
 p_2 \,  \Gamma^{456711} \big)   \,.}

With this choice,  \sampleKvec\ will generate Killing vectors parallel to
the brane, and a Killing vector, $L^\mu$ where $L^\mu {\del \over \del x^\mu} =
{\del \over \del x^8}$.   Using this last fact we can fix $p_2$ in \deformProj\
in terms of the Ansatz for the frame ${e^8}_\mu$, and then the function $p_1$ is fixed
by \quadreln.  Finally, the fact that \sampleKvec\ generates Killing vectors
parallel to the brane fixes the normalizations of the Killing spinors in
terms of the warp factor ({\it i.e.} the Ansatz for the frames ${e^j}_\mu,  j=1,2,3$) exactly as
it does for the harmonic distribution of frames.

We have thus {\it entirely determined} the Killing spinors in terms of
the metric Ansatz, and our task is to simply reconstruct everything else.
Indeed, we invert the usual perspective:  The Killing spinor
equation \gravvar\ is no longer a differential equation for $\epsilon$, it is an {\it
algebraic} equation for $F_{\mu\nu\rho\sigma}$.

\newsec{Solving the Ansatz: An example}

We are going to generalize the result of the \GowdigereUK\ by
considering a metric of the same general type, but with metric coefficients
that are arbitrary functions of two variables.  We thus take:
\eqn\metansatz{\eqalign{ds^2_{11}  ~=~ &  e^{2\, A_0} \,  ( -dt^2 +   dx_1^2 + 
dx_2^2) ~+~ e^{2\, A_1} \, du^2 ~+~ e^{2\, A_2} \, dv^2   ~+~ e^{2\, A_3} \, 
(\sigma_1^2  +\sigma_2^2)  \cr &  ~+~e^{2\, A_4} \, \sigma_3^2 ~+~ e^{2\, A_5} \, 
(\tau_1^2  +\tau_2^2)   ~+~e^{2\, A_6} \, \tau_3^2\,.}}
The $\sigma_j$ and $\tau_j$ are two independent sets of $SU(2)$ right-invariant
one-forms, which we will parametrize by Euler angles $\varphi_j$ and $\phi_j$
respectively.  They are normalized so that $d \sigma_1 = \sigma_2 \wedge \sigma_3$
and similarly for $\tau_j$.  The functions, $A_a =A_a(u,v), a=0,\dots,6$ will be taken to
be arbitrary functions of the coordinates $(u,v)$.  We have now made a rather special 
choice: the $x^{9}, x^{10}$ directions coincide with another $SU(2)$ symmetry action,
and so the combined choices of \metansatz\ and \gammastar\ are no longer so general.
We have, however made these choices so as to parallel, and generalize the results
of \GowdigereUK.  We now need to understand how the Killing spinors depend upon 
the coordinates along which symmetries act.  

First, the Poincar\'e invariance along the brane
means that  $\partial_\mu \epsilon =0$, $\mu=1,2,3$.
The metric Ansatz has a manifest  $SU(2)_\sigma \times U(1)_\sigma$
and $SU(2)_\tau \times U(1)_\tau$  symmetry, where the subscripts denote the
relevant $1$-forms in \metansatz.  Following \GowdigereUK, the Killing spinors that we 
seek are going to singlets under $SU(2)_\tau \times U(1)_\sigma$, and they 
will transform as ${\bf 2}_{\pm 1}$ under $SU(2)_\sigma \times U(1)_\tau$. The triviality of 
the $SU(2)_\tau$ action will mean that   $\partial_\mu \epsilon =0$, $\mu=9,10,11$, while
the non-trivial action of the $SU(2)$ $\cR$-symmetry in the $x^6, x^7, x^8$ directions  
enforces the appropriate doublet action of the isometry.    Define
\eqn\rotmat { \cR_{X}(\varphi) ~=~ \cos(\coeff{1}{2} \, \varphi ) \,
{\oneone} ~-~ \sin(\coeff{1}{2} \, \varphi ) \, \Gamma^{X} \,,} 
and let
\eqn\grot{ g~\equiv~  \cR_{67}(\varphi_3)\, \cR_{12378}(\varphi_1) \, \cR_{67}(\varphi_2)  \,.}
By construction one has:
\eqn\rtinv{ dg \, g^{-1} ~=~ - \coeff{1}{2}  \big( \sigma_1\,
 \Gamma^{12378} ~-~   \sigma_2\, \Gamma^{12368}~+~ \sigma_3\,  \Gamma^{67}\big)  \,.}
The Killing spinors are then of the form:
\eqn\Kspcoorddep{ \epsilon ~=~  g \, \hat \epsilon_0 \,, \qquad {\rm where} \quad \del_\mu 
\,\epsilon_0 ~=~ 0, \ \ \mu =1,2,3,6,7,8,9,10,11 \,. }

The slightly unusual feature is presence of the $\Gamma^{123}$ in \grot, which
means that the Clifford algebra for the $SU(2)_\cR$ is generated
by $\Gamma^{1236}$, $\Gamma^{1237}$ and $\Gamma^8$.  However, the 
presence of the extra $\Gamma^{123}$ factors is essential for these rotation
matrices to commute with the projector, $\Pi_0$, defined by \gammastar.  The slightly
unusual $SU(2)_\cR$ Clifford algebra is thus mandated by the deformation of
the fundamental projector. 

Using the Killing vector conditions on the Killing spinors one obtains:
\eqn\projfns{ \epsilon ~=~  e^{{1 \over 2}  A_0} \, g \,   \epsilon_0 \,, \qquad p_2 ~=~ 
\beta_0 \, e^{A_4 - A_0} \,, \quad p_1 ~=~ \sqrt{1 -p_2^2} \,.}
where $\beta_0$ is a constant, and $\epsilon_0$ is a constant spinor.  The parameter,
$\beta_0$ represents the strength of the deformation.

Rather than go through the general procedure outlined in the previous
section, we now focus the calculation towards generalizing the results
of \GowdigereUK.  To that end, we will require that the only non-zero
components of $F$ be:
\eqn\nonzeroF{F_{1234}\,, \  F_{1235}\,, \  F_{45811}\,, \  F_{67910}\,, \  
F_{4 67 11}\,, \  F_{56711}\,, \  F_{48910}\,, \  F_{58910}   \,.}
This is motivated by the general form of the solution in \GowdigereUK,
but here we will allow them to be general functions of $u$ and $v$.

Using this Ansatz one can easily see that the linear combinations:
\eqn\nicedeltas{\Gamma^1 \delta \psi_1~+~ \Gamma^6 \delta \psi_6~+~
\Gamma^8 \delta \psi_8~=~ 0 \,, \quad \Gamma^1 \delta \psi_1~+~ 
\Gamma^9 \delta \psi_9~+~ \Gamma^{11}  \delta \psi_{11}~=~ 0 \,,}
where all indices are frame indices, have the remarkable property that
the $F$-tensor terms cancel out.  The vanishing of these combinations of
gravitino variations lead to four conditions on the functions, $A_a$.  First,
one finds that  $A_0 + A_3 +A_4$  must be purely a function of $u$ and 
$A_0 + A_5 +A_6$ must be purely a function of $v$.
Since both  $u$ and $v$ are each arbitrary up to redefinitions
$u \to f_1(u)$, $v \to f_2(v)$,  for some arbitrary functions, $f_j$, 
we now fix this freedom completely by imposing:
\eqn\uvfix{ e^{A_0 + A_3 +A_4} ~=~ \coeff{1}{4} \, u^2  \,, \qquad
e^{A_0 + A_5 +A_6} ~=~ \coeff{1}{4} \, v^2 \,.}
Having done this, the other two conditions imposed by \nicedeltas\ 
mean that the functions $A_a$  can all be expressed in terms of three 
functions   $B_0, B_1, B_2$:  
\eqn\Arepl{\eqalign{ A_0 ~=~ & B_0 \,, \qquad  A_1 ~=~B_1 -  \coeff{1}{2} B_0 \,, \qquad 
A_2 ~=~ B_2 -  \coeff{1}{2} B_0 \,, \cr 
A_3 ~=~ & B_1 -  \coeff{1}{2} B_0 + \log\big(\coeff{1}{2} \,u \big)  \,, \qquad A_4 ~=~ 
- B_1 -  \coeff{1}{2} B_0 + \log\big(\coeff{1}{2} \,u \big)  \,,  \cr 
A_5 ~=~& - B_2 -  \coeff{1}{2} B_0 + \log\big(\coeff{1}{2} \,v \big)  \,, \qquad 
A_6 ~=~ B_2 -  \coeff{1}{2} B_0 + \log\big(\coeff{1}{2} \,v\big)    \,. }}
This means that the metric can be recast in the form:
\eqn\metansnew{ \eqalign{ds^2_{11}  ~=~ &  H^{-{2 \over 3}} \,  ( -dt^2 +   dx_1^2 + 
dx_2^2) ~+~ H^{{1\over 3}} \big[ V_1 \,\big( \, du^2 ~+~ \coeff{1}{4}\, u^2 (\sigma_1^2  +
 \sigma_2^2) \big)  \cr & ~+~V_1^{-1}  \,\big( \coeff{1}{4}\, u^2 \sigma_3^2  \big)  ~+~  
V_2 \,\big( \,  dv^2   ~+~  \coeff{1}{4}\, v^2 \,\tau_3^2 \,\big)     ~+~
V_2^{-1} \big( \coeff{1}{4}\, v^2 (\tau_1^2  +  \tau_2^2) \big) \,\Big]\,.}}
where
\eqn\metfndefs{ H ~\equiv~ e^{-3 \,B_0} \,, \qquad  V_1 ~\equiv~ e^{2 \,B_1}
 \,, \qquad  V_2 ~\equiv~ e^{2 \,B_2} \,.}

Using other parts of the gravitino variation, one finds a simple 
differential equation that relates $B_1$ and $B_2$.  This may be
conveniently written as:    
 \eqn\Bonetwo{ e^{2(B_1 + B_2)} ~=~ {1 \over 2\, v}\, {\del \over \del v}\,
\big( v^2 \,  e^{2(B_1 - B_2)} \big) \,.}
In other words, once one knows $(B_1 - B_2)$, one can use this 
to find $(B_1 + B_2)$.  One can also algebraically determine the components
of the Maxwell tensor, \nonzeroF.   The Bianchi identities then give equations
of motion for $B_0$ and  $(B_1 - B_2)$.  Rather than try to give an exhaustive
classification of the solutions of our more limited Ansatz, we will focus on
one particular family of solutions.

We take the $3$-form potential, $A^{(3)}$ to have the form:
\eqn\Aform{ A^{(3)} ~=~ q_1 \, dt \wedge dx_1 \wedge dx_2 ~+~ 
q_2 \, \sigma_1 \wedge \sigma_2 \wedge \tau_3  ~+~ 
q_3 \, \sigma_3 \wedge J \,,}
where the $q_j$ are arbitrary functions of $u$ and $v$, while the $2$-form
$J$ is given by:
\eqn\Jform{J ~=~\coeff{1}{2} \,  e^{2 \,B_2} \, v \, dv  \wedge  \tau_3  ~+~ 
\coeff{1}{4} \,  e^{- 2 \,B_2}  \, v^2 \, \tau_1 \wedge \tau_2 \,.}
This Ansatz for $A^{(3)}$ is, in fact deduced from relations \AdotK.  Indeed,
for suitably chosen $i,j$, the $2$-form components, $J_{\mu \nu}$,  can be extracted as 
part of $\Omega^{ij}_{\mu \nu}$.  One can use \gravvar\ with \Aform\ to
fix the $q_j$, or one simply apply \AdotK.  This leads immediately to 
\eqn\qonethree{q_1 ~=~\coeff{1}{2} \,  e^{3 \,B_0 + 2\, B_1}  \,, \qquad
q_3 ~=~-  {1 \over 2  \, \beta_0} \,  e^{- 2\, B_1}  \,,}
where $\beta_0$ is the constant in \projfns.  The only part of $A^{(3)}$
that is not fixed by \AdotK\ is $q_2$.  This {\it must} be obtained from \gravvar,
and one gets three equations for it.  Two of these are equivalent via \Bonetwo, 
and one of these two yields:
\eqn\qtwo{q_2 ~=~-  {1 \over 16  \, \beta_0} \,u^3 v^2 {\del \over \del u}
\Big( {1 \over u^2} \,  e^{ 2\, (B_1-B_2)}  \Big) \,.}
The third equation for $q_2$ is:
\eqn\qtwoother{{\del q_2  \over \del u} ~=~   {1 \over 8  \, \beta_0} \,u\,  v  \,
{\del  \over \del v} \big(  e^{ 4\, B_1 }  \big) \,.}
Write $e^{ 4\, B_1 } =e^{ 2\, (B_1-B_2) } e^{ 2\,  (B_1 + B_2)}$ and use 
\Bonetwo\ to eliminate $e^{ 2\, (B_1+B_2) } $, and one gets:
\eqn\qtwointer{ {\del q_2  \over \del u} ~=~   {1 \over 8  \, \beta_0} \,u\,  v  \,
{\del  \over \del v} \Big( {1 \over 2\, v}\,  e^{ 2\, (B_1-B_2) } {\del \over \del v}\,
\big( v^2 \,  e^{2(B_1 - B_2)} \big) \Big) \,.}
Comparing this with \qtwo, one sees that all the conditions on $q_2$
are satisfied if and only if:
\eqn\DEfirst{ {1 \over u^3}\, {\del \over \del u}\Big( \,u^3 {\del \over \del u}
\Big( {1 \over u^2} \,  e^{ 2\, (B_1-B_2)}  \Big) \Big) ~=~ -  {1 \over v}  \,
{\del  \over \del v} \Big(  e^{ 2\, (B_1-B_2) }\, {1 \over  v}\,  {\del \over \del v}\,
\Big( {v^2  \over u^2}  \,  e^{2(B_1 - B_2)} \Big) \Big) \,,}
which is precisely \basicPDE.    Finally, the complete solution of \gravvar\  determines
$B_0$ algebraically:
\eqn\Bzero{e^{ - 3B_0 } ~=~  {4 \over  \,   \beta_0\, u^2} \,
 \big( e^{ 2\, B_1} ~-~ e^{ -2\, B_1} \big) \,.}

Thus, a complete solution to \gravvar\ is obtained by solving \DEfirst.  From the
solution one can obtain $B_1$ and $B_2$ independently using \Bonetwo.
Then $B_0$ is obtained from \Bzero.  The functions $q_j$ are then given by
\qonethree\ and \qtwo.  This completely fixes the solution.

One can also check that this solution solves all the equations of
motion \eqnmot, and not merely a subset of them.    

In deriving the solution above we did not try to find the most general solution: 
We imposed some special Ans\"atze and discarded some constants of integration.
Our purpose was to illustrate the power of the general method.  It is
interesting to note that every function, save one ($q_2$) was fixed algebraically
in terms of the metric coefficients, and that it was the non-algebraic function
that gave rise to the only differential equation that needs to be satisfied.

To conclude this section, we obtain some special solutions to the
foregoing procedure.   First,  note that we can easily recover the result of \GowdigereUK\  
via the change of coordinates:
\eqn\coordchg{u ~=~ {\rho \over \sqrt{\sinh(2\,\chi)} } \, \cos\theta\,, 
\qquad v ~=~ {1 \over \sqrt{\sinh(2\,\chi)} } \sin\theta \,,}
where the quantities $\rho, \chi, \theta$ are defined in \GowdigereUK.
In particular, $e^{ 2\, (B_1-B_2)} = \cosh(2\, \chi)$ is a solution to
\DEfirst.   The form of  $A^{(3)}$  presented here is not exactly the 
same as that of \GowdigereUK, but it is gauge equivalent.

It is also interesting to find the ``separable'' solutions in which one
seeks solutions of \DEfirst\ with  $e^{ 2\, (B_1-B_2)} =h_1(u) h_2(v)$ for 
some functions $h_1$ and  $h_2$.  Obviously separation of variables
does not work in general due to the non-linearity, but we do find
a solution to \DEfirst\ and \Bonetwo\ with:
\eqn\sepsol{e^{ 2\,  B_1} ~=~ \mu  \,(1 + b \, u^2)  \,, \qquad 
e^{ 2\,  B_2} ~=~  \Big(1 \pm  \Big( {a \over  v} \Big)^4  \Big)^{-{1 \over 2}} \,,} 
for some constants, $a, b$ and $\mu$.  One can then go on to
find the expressions for $e^{ 3\,  B_0}$ and the $q_j$, and thus obtain
a three-parameter family of solutions.   The asymptotics at 
large distances are determined   by  $b$ and $\mu$:
If $b = 0$ then the metric in the $u$-direction that of $\IR^4$, but with a stretched
Hopf fiber over the $S^3$ at fixed $v$.  The amount of stretching is given by
$\mu$.   If $b \ne 0$ then the metric is that of  $\IR^3 \times S^1$.
If one wants the metric to asymptote to that of $AdS_4 \times S^7$ then one must set 
$\mu = 1$ and $b=0$.  However, this limit must be taken carefully:
to get a finite result for $e^{ 3\,  B_0}$  from \Bzero\ one has to set 
$\mu = 1 + \epsilon^2$ and $\beta_0 = \alpha\, \epsilon$, for some constant, $\alpha$, 
and then take the  limit as $\epsilon \to 0$.  Thus the projector, $\Pi_0$, becomes the 
standard one parallel to the $M_2$ branes (\ie\ it has $p_1 =1$ and $p_2=0$).  
One then finds a solution in which the $q_j$'s are non-zero, but all the ``internal''
components are {\it pure gauge}: That is, the only non-zero component of the 
field strength is $F_{1234}$.  One also finds that $H = e^{-3 \, B_0} = 
{4 \over \alpha^2 \, u^2}$.   The solution is therefore that of $M2$ branes
uniformly spread over the $\IR^4$ in the $(5,9,10,11)$ directions.   There is, however,
a small variation from the usual harmonic-brane story:  For $a \ne 0$,
the metric in  direction of brane spreading is not flat, but is the Eguchi-Hanson metric.
This can be seen more directly by taking the $+$-sign choice in \sepsol, and changing 
variables to $w \equiv  (v^4 + a^4)^{1/4}$.
 
While we have constructed as set of rather special solutions, we would
like to stress that, as pointed out in the
introduction, there is an infinite  family of solutions to \DEfirst.  These
solutions can at least be generated by perturbation theory, and presumably 
correspond to a general, rotationally symmetric, $v$-dependent 
distributions of $M2$-branes in the $x^5, x^9, x^{10}, x^{11}$ directions.
It is also important to note that in \GowdigereUK\  the ``master function''  is
given by  $e^{ 2\, (B_1 -  B_2)}= \cosh(2\, \chi)$ and that $\chi$  is the gauged
supergravity scalar dual to the fermion mass.   Thus  the entire flow is
determined by the flow of this mass term.

\newsec{Some comments on the geometry of the transverse eight-manifold}

The metric \metansnew\ has quite a number of interesting geometric features
that we will expand upon in \GNW.  Consider the metric, $ds_8^2$, in the square
brackets:  
\eqn\eightmet{ \eqalign{ds^2_{8}  ~=~   ds_{4}^2 ~+~   d\hat s_{4}^2  ~\equiv~&
  V_1 \,\big( \, du^2 ~+~ \coeff{1}{4}\, u^2 (\sigma_1^2  +
 \sigma_2^2) \big)  ~+~V_1^{-1}  \,\big( \coeff{1}{4}\, u^2 \sigma_3^2  \big)  \cr & ~+~  
V_2 \,\big( \,  dv^2   ~+~  \coeff{1}{4}\, v^2 \,\tau_3^2 \,\big)     ~+~
V_2^{-1} \big( \coeff{1}{4}\, v^2 (\tau_1^2  +  \tau_2^2) \big)  \,,}}
where the split into $ds_{4}^2$ and $d\hat s_{4}^2$ corresponds to the split
into $u, \sigma_j$ and $v, \tau_j$ respectively.

The form of $ds_{4}^2$ is very reminiscent the  Gibbons-Hawking ALE metrics \GibbonsZT,
and to get it to  the same form one simply needs to make a change
of variable to remove a conical singularity at $u=0$.  That is, set 
$u = \sqrt{w}$, and define $\widetilde V_1 = u^{-2} V_1$, and one then finds:
\eqn\dsfoursimp{ ds_{4}^2  ~=~ \coeff{1}{4}\,  \big[ \, \widetilde  V_1 \,\big( \, dw^2 ~+~  
 w^2  (\sigma_1^2  + \sigma_2^2) \big)  ~+~ \widetilde V_1^{-1}  \,\big( w^2 
\sigma_3^2  \big)\, \big]}
The metric in parentheses is now precisely that of a  flat $\IR^3$, and there is the
$S^1$ fibration over this $\IR^3$.  The only difference with the Gibbons-Hawking
form is that the function, $\widetilde V_1$, does not appear to be harmonic 
on the $\cR^3$.  The function $V_2^{-1} \widetilde V_1$ satisfies \DEfirst, whose 
left-hand side involves:  
$$
{1 \over w^2}\, {\del \over \del w}\Big( \,w^2 {\del \over \del w}
\Big( V_2^{-1} \,\widetilde V_1 \Big) \Big) \,,
$$
which is the Laplacian on $\IR^3$.  As yet, we do not know if \DEfirst\ can then
be translated into some interesting generalization of the harmonic condition on 
$ \widetilde V_1$. 

The appearance of this $S^1$ fibration over a flat $\IR^3$ is a significant new feature of 
our formulation here.    This four-dimensional space is where the $\cR$-symmetry acts,
and it lies transverse to both the branes and to their moduli space.  We believe
that the geometry in this direction will always have the same form, independent of how
the branes spread.   Indeed, the same geometry appears in the $\cN=2^*$ flows
(with eight supersymmetries) in IIB supergravity \PWIIB.
The description of  $ds_4^2$ given here is much simpler, and certainly 
more intuitive than that given in \GowdigereUK.  This is because the 
coordinates \coordchg\ are more naturally adapted to the decomposition of the geometry
of $ds_8^2$ transverse and parallel to the moduli space of the branes.  In particular, 
the selection of the coordinates, $(u,v)$, is directly linked to the choice of the projector, 
$\Pi_2$, in \Projone.   

The geometry of  $d\hat s_{4}^2$ is equally tantalizing:  We know that at $u=0$  it is
hyper-K\"ahler with three harmonic $2$-forms.  It turns out that these have natural
extensions to the whole space.   To see this, we first
introduce the usual set of Euler angles to define the left-invariant
$1$-forms, $\tau_j$:
\eqn\oneforms{\eqalign{\tau_1 ~\equiv~&  \cos \phi_3\, d\phi_1 ~+~ 
\sin\phi_3\, \sin\phi_1\, d \phi_2 \,, \cr
\tau_2 ~\equiv ~&  \sin\phi_3\, d\phi_1 ~-~ 
\cos\phi_3\, \sin\phi_1\, d \phi_2 \,, \cr
\tau_3 ~\equiv ~&  \cos\phi_1\, d\phi_2 ~+~   d \phi_3  \,,}}
The $2$-form, $J$, in \Jform\ {\it is} a complex structure for $d\hat s_{4}^2$, but it is not 
K\"ahler.  The corresponding complex coordinates are are simply:
\eqn\cplxcoords{\zeta_1 ~\equiv~ v \, \cos(\coeff{1}{2} \phi_1) \, e^{{i \over 2}\,
(\phi_2 + \phi_3)} \,, \qquad 
\zeta_2 ~\equiv~ v \, \sin (\coeff{1}{2} \phi_1) \, e^{- {i \over 2}\,
(\phi_2 -\phi_3)} \,.}
The other two $2$-forms are given by the real and imaginary parts of:
\eqn\specforms{\Omega  ~\equiv~\coeff{1}{2} \,  ( dv ~+~ \coeff{i}{2}    \,  v\, \tau_3) ~ 
\wedge ~ v  \,( \tau_1 + i\, \tau_2) ~= ~  d \zeta_1  ~ \wedge ~ d \zeta_2 \,.}
These are global, harmonic $2$-forms on the whole of $ds_8^2$.

 It is relatively straightforward to check that the differential forms, $J$ and $\Omega$
 play a significant role in the underlying $G$-structure of our solution.

\newsec{Final comments}

There are now many interesting flow solutions that have been constructed
using  gauged supergravity theories.  The advantage of such an approach has been
the simplicity of the equations of motion in the lower dimensional theory.  However,
if one reviews the solutions in four and five dimensions for which the $M$-theory
or $IIB$ ``uplift''  is known, then one encounters what appear to be some extremely 
involved solutions.    Even if there are high levels of supersymmetry, these solutions 
are still very complicated and they have metric and tensor gauge fields that appear to defy 
simple classification (see, for example, \refs{\PilchEJ\PilchUE\PilchFU\KhavaevYG{--}
\CorradoNV,\GowdigereUK}).  One of the important messages of this paper is
that the complexity of these previously known solutions lies in the fact that 
one is focusing on the wrong object:  The solutions are very simple if one focuses
one the Killing spinors.  The contrast between the solution presented here and
that described in \GowdigereUK\ illustrates this graphically, and the solution presented
here is considerably more general.

Our strategy for finding solutions is to make a very general Ansatz for the metric,
based upon the Poincar\'e and $\cR$-symmetries, and upon the presence
of a moduli space for the brane.  We then make an 
Ansatz for the Killing spinors by defining projectors {\it algebraically} in
terms of the metric Ansatz.   There are several crucial inputs into the
projector Ansatz:  (i)  Deforming the canonical ``Dirichlet projector'' parallel
to the branes,  (ii) The $\cR$-symmetries, (iii) Poincar\'e invariance
along the brane,  (iv)  Spinor bilinears generate Killing vectors, 
and (v)  Projectors on the moduli space of the branes.    Once an projector
Ansatz is made, any arbitrary functions are fixed (algebraically) in terms of the 
metric Ansatz because of the Killing vector condition on spinor bilinears.
One then uses the supersymmetry variations of the fermions to fix
(algebraically) the field strengths of the tensor gauge fields.  This system of
equations is generically  highly over-determined and so it  also gives rise to first-order 
differential equations for some of the metric coefficients.  The complete set of equations
for the solution are then given by the Bianchi identities of the field strengths.
One should then verify that the solution to the supersymmetry variations
does indeed satisfy the equations of motion.  However,
we believe that the  equations coming from the supersymmetry variations
will generically be sufficient to solve all the equations
of motion except in highly specialized, algebraically simple solutions like 
the pure ``harmonic brane'' distributions.  In these special circumstances, 
commutators of supersymmetry variations generate combinations of
equations of motion in which there are non-trivial cancellations, and the
result is a only a subset of the constraints on the metric functions.  This is
the one circumstance where complexity helps:  If the solution is sufficiently
complex, such cancellations do not happen, and solving the supersymmetry variations
will capture everything.

The result presented here fits very nicely with what one expects from the
RG flow of the theory on the brane.   Specifically, the flow is driven by
the fermion mass parameter because this is the leading term in the field
theory Langrangian, with the bosonic mass term fixed by supersymmetry.
The complete holographic flow solution presented here is generated from
a single function, $g = e^{2(B_1 -B_2)}$,  that is obtained as a 
solution to the differential equation \basicPDE.  This leading part of this 
function is also dual to the fermion mass term, and the entire solution is being 
determined essentially by the flow of this one term.   We also suspect that this observation 
lies at the root of the success of the ``Algebraic Killing Spinor'' Ansatz.   That is,  from
the field theory perspective  these flows are very simple because they are
driven by a single relevant operator in the (supersymmetric) Langrangian
combined with a deformation along the Coulomb branch.  In holography, the
Coulomb branch is at the root of the ``harmonic rule,'' while the fermion mass
lies at the root of the deformation of the standard Dirichlet projector.   All we
have really done here is to try to combine these ideas.  

We suspect that the ideas presented in this paper will have many applications
in generating not just holographic flow solutions, but also in obtaining 
more general supersymmetric compactifications with non-trivial fluxes.  There
are some obvious  questions about using this to find more general classes
of flow solution.  There are open issues about classifying the possible
supersymmetric deformations of the Dirichlet projectors, and then there
are  questions  about the mathematical structures that underlie the more general
classes of solution.  What we have done here is to use projectors to
define the supersymmetries as a special sub-bundle of the spin bundle.
One should be able to find an intrinsic classification of this special sub-bundle,
and since the projectors are algebraic in the metric Ansatz one should probably try
to classify the total space of this spinor sub-bundle.
Then there is an issue of the underlying $G$-structures and how they 
fit into the scheme presented here.   Finally, there is an interesting question
about how boundary conformal field theory relates to the deformation 
of the Dirichlet projectors.

\bigskip
\leftline{\bf Acknowledgements}

NW  would like to thank J.~Gomis for helpful conversations.  
This work was supported in part by funds
provided by the DOE under grant number DE-FG03-84ER-40168.

\listrefs
\vfill
\eject
\end